\documentstyle[12pt]{article}
\begin{document}
\begin{flushright}
hep-th/0510029\\
SNB/October/2005
\end{flushright}
\vskip 2cm
\begin{center}
{\bf \Large {Superfield Approach To Exact And Unique Nilpotent Symmetries }}

\vskip 2.5cm

{\bf R.P.Malik}
\footnote{ Talk delivered in the International Workshop on
``Supersymmetries and Quantum Symmetries'' (27-31 July 2005)
held at the Bogoliubov Laboratory of Theoretical Physics,
JINR, Dubna (Moscow), Russia. }\\
{\it S. N. Bose National Centre for Basic Sciences,} \\
{\it Block-JD, Sector-III, Salt Lake, Calcutta-700 098, India} \\
{\bf  E-mail: malik@boson.bose.res.in  }\\

\vskip 2.5cm

\end{center}

\noindent
{\bf Abstract}:
In the framework of {\it usual} superfield approach,
we derive the exact local, covariant,
continuous and off-shell nilpotent Becchi-Rouet-Stora-Tyutin (BRST)
and anti-BRST symmetry
transformations for the $U(1)$ gauge field
($A_\mu$) and the (anti-)ghost fields ($(\bar C)C$) of the Lagrangian
density of a four ($3 + 1$)-dimensional QED by exploiting the
horizontality condition defined on the
six ($4, 2)$-dimensional supermanifold. The long-standing problem of
the exact derivation of the above nilpotent
symmetry transformations for the matter (Dirac)
fields ($\bar \psi, \psi$), in the framework of superfield formulation, is
resolved by a new restriction on
the $(4, 2)$-dimensional supermanifold. This new
gauge invariant restriction on the supermanifold, due to
the {\it augmented} superfield formalism, owes its
origin to the (super) covariant derivatives.
The geometrical interpretations for all the above off-shell nilpotent
transformations are
provided in the framework of {\it augmented} superfield formalism.
\baselineskip=16pt

\vskip 1cm



\newpage

\noindent
The current year 2005 has been declared as the ``world year of physics''
to mark the 100th anniversary
of the epoch-making discoveries made by Einstein in his miraculous year
1905. The year 2005 has also been a landmark year for the researchers,
working in the realm of
Becchi-Rouet-Stora-Tyutin (BRST) formalism, because
it has celebrated the 30th birth anniversary of the discovery of BRST
symmetries in the context of gauge theories [1,2]. This formalism, during
its three decades of existence,
has found applications in some of the frontier areas of research like
topological field theories [3,4] and string field theories [5].

The key ideas of the BRST formalism
have deep connections with the mathematics
of differential geometry and (theoretical) physics of
gauge theories as well as
supersymmetries. One of its intuitive connections  is with
supersymmetry through the {\it usual} superfield formulation [6] which
provides the
geometrical interpretations for the nilpotent ($Q_{(a)b}^2 = 0$)
and anticommuting ($Q_b Q_{ab} + Q_{ab} Q_b = 0$) (anti-)BRST charges
($Q_{(a)b}$) in a beautiful manner. There exist, however, some
long-standing problems in this domain of research
which have defied their resolutions during the last 25 years.
In our presentation, we shall
touch upon one such long-standing problem (connected with the superfield
approach to BRST formalism) and provide its resolution by exploiting
the importance of {\it gauge invariance}.

Under the usual superfield approach [6], a $D$-dimensional Abelian
gauge theory (endowed with the first-class constraints in the
language of Dirac's prescription [7,8]) is considered on a $(D,
2)$-dimensional supermanifold parameterized by $D$-number of
spacetime (even) co-ordinates $x^\mu$ ($\mu = 0, 1, 2, 3....D-1$)
and a couple of (odd) Grassmannian variables $\theta$ and
$\bar\theta$ (with $\theta^2 = \bar\theta^2 = 0, \theta\bar\theta
+ \bar\theta \theta = 0$). In general, the $(p + 1)$-form super
curvature $\tilde F^{(p + 1)} = \tilde d \tilde A^{(p)}$,
constructed from the super exterior derivative $\tilde d$ (with
$\tilde d^2 = 0$) and the super $p$-form connection $\tilde
A^{(p)}$ (corresponding to a $p$-form ($p = 1, 2...$) Abelian
gauge theory) is restricted to be flat along the Grassmannian
directions of the $(D, 2)$-dimensional supermanifold due to the
so-called horizontality condition \footnote{ Nakanishi and Ojima
call it the ``soul-flatness'' condition [9]. For the 1-form
non-Abelian gauge theory, $\tilde F^{(2)} = \tilde d \tilde
A^{(1)} + \tilde A^{(1)} \wedge \tilde A^{(1)}$ and $F^{(2)} = d
A^{(1)} + A^{(1)} \wedge A^{(1)}$ in the horizontality condition
$\tilde F^{(2)} = F^{(2)}$ [6].}. Mathematically, this condition
implies $\tilde F^{(p + 1)} = F^{(p + 1)}$ where $F^{(p + 1)} = d
A^{(p)}$ is the $(p + 1)$-form curvature defined on the ordinary
$D$-dimensional manifold through the ordinary exterior derivative
$d= dx^\mu \partial_\mu $ (with $d^2 = 0$) and ordinary $p$-form
Abelian connection $A^{(p)} = \frac{1}{p!} \;[dx^{\mu_1} \wedge
dx^{\mu_2}...\wedge dx^{\mu_p}]\; A_{\mu_1\mu_2.....\mu_p}$.

The above horizontality condition on the six $(4, 2)$-dimensional
supermanifold leads to the
derivation of the nilpotent (anti-)BRST symmetry transformations for
the gauge- and (anti-)ghost fields of the (anti-)BRST invariant Lagrangian
density of a given four $(3 + 1)$-dimensional (4D) 1- and 2-form
(non-)Abelian gauge theories [6].
However, it does not shed any light on the nilpotent (anti-)BRST symmetry
transformations that are associated with the matter (Dirac) fields
of the interacting 1-form (non-)Abelian gauge theories where there is a
coupling between the gauge field and the matter conserved current,
constructed by the Dirac fields. This
issue (i.e. the derivation of the nilpotent transformations for matter fields)
has been a long-standing problem in the superfield
approach to BRST formalism.

In a recent set of papers
\footnote{The author is grateful to the {\bf ``Dubna School''} for his
training in supersymmetry and related topics.}
[10-13], the usual superfield formalism has been
consistently extended by invoking the additional restrictions on the six
$(4, 2)$-dimensional supermanifold that are complimentary to
the horizontality condition [6]. These additional restrictions
on the supermanifold are the equality of (i) the conserved (super)
matter current [10,11] (as well as other conserved quantities [11]), and
(ii) the gauge invariant quantities owing their origin to the
(super) covariant derivatives on the (super) matter fields [12,13].

The former set of restrictions [10,11]
lead to the consistent derivation of the nilpotent symmetry transformations
for the  matter fields. On the other hand, the latter restrictions
[12,13] lead to the exact and unique derivation of the nilpotent
symmetry transformations for the matter fields. We christen these
extended versions of the usual superfield approach to BRST formalism as
the augmented superfield formalism.
Both types of extensions
have their own merits and advantages. Any further (consistent)
extension of the usual superfield approach would be
a welcome sign for the future of this area of research.

In our presentation, we {\it first} focus on the strength of the horizontality
condition in the exact and unique
derivation of the nilpotent symmetry transformations
for the gauge and (anti-)ghost fields of a 4D interacting $U(1)$ gauge
theory with the Dirac fields. This interacting
Abelian system has been taken into consideration
{\it only} for the sake of simplicity. The ideas, proposed
in our presentation, can be
generalized to a non-Abelian interacting gauge theory in a straightforward
manner. Second, we concentrate on the consistent
derivation of the nilpotent transformations for the matter (Dirac) fields
by exploiting the equality of the conserved matter (super) current
on the six $(4, 2)$-dimensional supermanifold. Finally,
we obtain the exact and unique nilpotent symmetry
transformations for the Dirac fields by exploiting the equality of the
gauge invariant quantity on the above supermanifold
that owes its origin to the (super) covariant
derivatives on the (super) Dirac fields.

Let us begin with the (anti-)BRST invariant Lagrangian density ${\cal L}_{b}$
for the {\it interacting} four ($3 + 1)$-dimensional $U(1)$ gauge theory
in the Feynman gauge [14]
$$
\begin{array}{lcl}
{\cal L}_{b} = - \frac{1}{4}\; F^{\mu\nu} F_{\mu\nu}
+ \bar \psi \;(i \gamma^\mu D_\mu - m)\; \psi + B \;(\partial \cdot A)
+ \frac{1}{2}\; B^2
- i \;\partial_{\mu} \bar C \partial^\mu C,
\end{array} \eqno(1)
$$
where $F_{\mu\nu} = \partial_\mu A_\nu - \partial_\nu A_\mu$ is the
antisymmetric field
strength tensor for the $U(1)$
Abelian gauge theory that is derived from the 2-form
$d A^{(1)} = \frac{1}{2} (dx^\mu \wedge dx^\nu) F_{\mu\nu}$
\footnote{We adopt here the conventions and notations such that the 4D flat
Minkowski metric is: $\eta_{\mu\nu} =$ diag $(+1, -1, -1, -1)$ and $\Box =
\eta^{\mu\nu} \partial_{\mu} \partial_{\nu} = (\partial_{0})^2
- (\partial_i)^2, F_{0i}
= E_i = \partial_{0} A_{i} - \partial_{i} A_{0} = F^{i0}, F_{ij} =
\epsilon_{ijk} B_k, B_i = (1/2) \epsilon_{ijk} F_{jk},
D_{\mu} \psi = \partial_{\mu} \psi + i e
A_{\mu} \psi$ where $\epsilon_{ijk}$ is the 3D
totally antisymmetric Levi-Civita tensor and electric and magnetic
fields are $E_i$ and $B_i$, respectively. In equation (1),
 $\gamma$'s are the usual
$4 \times 4$ Dirac matrices. Furthermore, the Greek indices: $\mu, \nu, \rho...
= 0, 1, 2, 3$ in (1), correspond to the spacetime directions
and Latin indices $i, j, k...= 1, 2, 3$ stand {\it only} for the space
directions on the 4D spacetime manifold.}.
As is evident, the latter
is  constructed by the application of the exterior derivative
$d = dx^\mu \partial_\mu$ (with $d^2 = 0)$ on the 1-form
$A^{(1)} = dx^\mu A_\mu$ which defines the Abelian vector potential $A_\mu$.
The gauge-fixing term $(\partial \cdot A)$ is derived through the operation
of the co-exterior derivative $\delta$
(with $\delta = - * d *, \delta^2 = 0$) on the
one-form $A^{(1)}$ (i.e. $\delta A^{(1)} = - * d * A = (\partial \cdot A)$)
where $*$ is the Hodge duality operation. The fermionic
Dirac fields $(\psi, \bar \psi)$, with the mass $m$ and charge $e$, couple
to the $U(1)$ gauge field $A_\mu$ (i.e. $ - e \bar \psi \gamma^\mu A_\mu \psi$) through the conserved current
$J_\mu = \bar \psi \gamma_\mu \psi$. The
anticommuting ($ C \bar C + \bar C C = 0, C^2 = \bar C^2 = 0,
C \psi + \psi C = 0$ etc.) (anti-)ghost fields $(\bar C)C$ are required to
maintain the unitarity and ``quantum'' gauge (i.e. BRST) invariance together
at any arbitrary order of perturbation theory for a given physical process
\footnote{ The full strength of the (anti-)ghost fields turns up in the
discussion of the unitarity and ``quantum'' gauge (i.e. BRST) invariance
for the perturbative computations in the realm of non-Abelian gauge theory
where, for each loop diagram of the gauge (gluon) fields corresponding to
a physical process, a loop diagram consisting of
{\it only} the (anti-)ghost fields is required to exist as its counterpart
(see, e.g., [15] for details).}.
The Nakanishi-Lautrup auxiliary field $B$ is
required to linearize the quadratic gauge-fixing term
$-\frac{1}{2} (\partial\cdot A)^2$, present
in the Lagrangian density (1), in a subtle way.

The above Lagrangian density
(1) respects the following off-shell nilpotent
$(s_{(a)b}^2 = 0)$ and anticommuting ($s_b s_{ab} + s_{ab} s_b = 0$)
(anti-)BRST ($s_{(a)b}$)
\footnote{We adopt here the notations and conventions followed in [14].
In fact, in its full glory, a nilpotent ($\delta_{B}^2 = 0$)
BRST transformation $\delta_{B}$ is equivalent to the product of an
anticommuting ($\eta C = - C \eta, \eta \bar C = - \bar C\eta,
\eta \psi = - \psi \eta, \eta \bar \psi = - \bar \psi \eta$ etc.)
spacetime independent parameter $\eta$ and $s_{b}$
(i.e. $\delta_{B} = \eta \; s_{b}$) where $s_{b}^2 = 0$.}
symmetry transformations [14]
$$
\begin{array}{lcl}
s_{b} A_{\mu} &=& \partial_{\mu} C, \qquad
s_{b} C = 0, \qquad
s_{b} \bar C = i B,  \qquad s_b \psi = - i e C \psi, \nonumber\\
s_b \bar \psi &=& - i e \bar \psi C,
\qquad   s_{b} B = 0, \quad
\;s_{b} F_{\mu\nu} = 0, \quad s_b (\partial \cdot A) = \Box C, \nonumber\\
s_{ab} A_{\mu} &=& \partial_{\mu} \bar C, \qquad
s_{ab} \bar C = 0, \qquad
s_{ab} C = - i B,  \qquad s_{ab} \psi = - i e \bar C \psi, \nonumber\\
s_{ab} \bar \psi &=& - i e \bar \psi \bar C,
\qquad  s_{ab} B = 0, \quad
\;s_{ab} F_{\mu\nu} = 0, \quad s_{ab} (\partial \cdot A) = \Box \bar C.
\end{array}\eqno(2)
$$
The noteworthy points, at this stage, are (i) under the nilpotent (anti-)BRST
transformations, it is the kinetic energy term
(more precisely $F_{\mu\nu}$ itself) that remains invariant. (ii)
The electric and magnetic fields $E_i$ and $B_i$ (that are components of
$F_{\mu\nu}$) owe their origin to the operation of
cohomological operator $d$ on the one-form $A^{(1)}$. (iii)
The symmetry transformations in (2) are generated  by
the local, conserved and nilpotent charges $Q_{(a)b}$. This
statement, for the local generic field $\Sigma (x)$,
can be succinctly expressed as
$$
\begin{array}{lcl}
s_{r}\; \Sigma (x) = - i\;
\bigl [\; \Sigma (x),  Q_r\; \bigr ]_{\pm}, \qquad\;\;\;
r = b, ab,
\end{array} \eqno(3)
$$
where
$\Sigma (x) = A_\mu (x), C (x), \bar C (x), \psi (x), \bar \psi (x), B (x)$
and the $(+)-$ signs,
as the subscripts on the square bracket, correspond to the (anti-)commutators
for the generic local field $\Sigma (x)$
(of the Lagrangian density (1)) being (fermionic)bosonic in nature.

To derive the above anticommuting and
nilpotent transformations $s_{(a)b}$ for the bosonic
$U(1)$ gauge field $A_\mu$
and the fermionic
(anti-)ghost fields $(\bar C)C$, we exploit the {\it usual}
superfield formalism, endowed with the horizontality restriction on
a six ($4, 2$)-dimensional supermanifold. This supermanifold is
parametrized by the superspace coordinates $Z^M = (x^\mu, \theta, \bar \theta)$
where $x^\mu\; (\mu = 0, 1, 2, 3)$ are a set of four
even (bosonic) spacetime coordinates
and fermionic $\theta$ and
 $\bar \theta$ are a set of two odd (Grassmannian) coordinates.
One can define a super 1-form $\tilde A^{(1)} = dZ^M \tilde A_M$ where
the supervector superfield $\tilde A_M$ (with
$\tilde A_M = (B_{\mu} (x, \theta, \bar \theta),
\;{\cal F} (x, \theta, \bar \theta),
\;\bar {\cal F} (x, \theta, \bar \theta))$ has the component multiplet
superfields $B_\mu, {\cal F}, \bar {\cal F}$.
These component superfields can be expanded
in terms of the basic fields ($A_\mu, C, \bar C$), auxiliary field
($B$) of the Lagrangian density (1) and some extra secondary fields, as [6]
$$
\begin{array}{lcl}
B_{\mu} (x, \theta, \bar \theta) &=& A_{\mu} (x)
+ \theta\; \bar R_{\mu} (x) + \bar \theta\; R_{\mu} (x)
+ i \;\theta \;\bar \theta S_{\mu} (x), \nonumber\\
{\cal F} (x, \theta, \bar \theta) &=& C (x)
+ i\; \theta \bar B (x)
+ i \;\bar \theta\; {\cal B} (x)
+ i\; \theta\; \bar \theta \;s (x), \nonumber\\
\bar {\cal F} (x, \theta, \bar \theta) &=& \bar C (x)
+ i \;\theta\;\bar {\cal B} (x) + i\; \bar \theta \;B (x)
+ i \;\theta \;\bar \theta \;\bar s (x).
\end{array} \eqno(4)
$$
It is straightforward to note that the local
fields $ R_{\mu} (x), \bar R_{\mu} (x),
C (x), \bar C (x), s (x), \bar s (x)$ are fermionic (anticommuting)
in nature and their number matches with
the bosonic (commuting) local fields $A_{\mu} (x), S_{\mu} (x),
{\cal B} (x), \bar {\cal B} (x), B (x), \bar B (x)$ in (4).

All the secondary fields will be expressed
in terms of basic fields ($A_\mu, C, \bar C$) and the auxiliary field ($B$)
due to the restrictions emerging from the application
of horizontality condition. The explicit forms
of $\tilde F^{(2)}$ and $F^{(2)}$, in the horizontality restriction, are:
$$
\begin{array}{lcl}
\tilde F^{(2)} = F^{(2)},\; \;
\tilde F^{(2)}
= \tilde d \tilde A^{(1)} =  \frac{1}{2} (d Z^M \wedge d Z^N)
\tilde F_{MN}, \; \;F^{(2)} =
d A^{(1)} = \frac{1}{2} (dx^\mu \wedge dx^\nu) F_{\mu\nu}.
\end{array} \eqno(5)
$$
The super exterior derivative $\tilde d$ and
the connection super one-form $\tilde A^{(1)}$, in (5), are
$$
\begin{array}{lcl}
\tilde d &=& \;d Z^M \;\partial_{M} = d x^\mu\; \partial_\mu\;
+ \;d \theta \;\partial_{\theta}\; + \;d \bar \theta \;\partial_{\bar \theta},
\nonumber\\
\tilde A^{(1)} &=&
d Z^M\; \tilde A_{M} = d x^\mu \;B_{\mu} (x , \theta, \bar \theta)
+ d \theta\; \bar {\cal F} (x, \theta, \bar \theta) + d \bar \theta\;
{\cal F} ( x, \theta, \bar \theta).
\end{array}\eqno(6)
$$
Mathematically, the above condition (5) implies
the ``flatness'' of all the components of the (anti-)symmetric
super curvature tensor $\tilde F_{MN}$ that are directed along the
 $\theta$ and/or $\bar \theta$ directions of the supermanifold.
Ultimately, the soul-flatness (horizontality) condition
($\tilde d \tilde A^{(1)} = d A^{(1)}$) of the equation (5) (with
$\tilde F^{(2)} = F^{(2)}$), yields
\footnote{In the explicit computation of $\tilde d \tilde A^{(1)}$,
we have taken into account $dx^\mu \wedge dx^\nu = - dx^\nu \wedge dx^\mu,
dx^\mu \wedge d\theta = - d \theta \wedge dx^\mu, d\theta \wedge d \bar\theta
= d \bar\theta \wedge d\theta$, etc., that emerge from the requirement
of the nilpotency of $\tilde d$ (i.e. $\tilde d^2 = 0$).}
$$
\begin{array}{lcl}
R_{\mu} \;(x) &=& \partial_{\mu}\; C(x), \qquad
\bar R_{\mu}\; (x) = \partial_{\mu}\;
\bar C (x), \qquad \;s\; (x) = \bar s\; (x) = 0,
\nonumber\\
S_{\mu}\; (x) &=& \partial_{\mu} B\; (x)
\qquad
B\; (x) + \bar B \;(x) = 0, \qquad
{\cal B}\; (x)  = \bar {\cal B} (x) = 0.
\end{array} \eqno(7)
$$
The insertion of all the above values in expansion (4) leads to
the derivation of (anti-)BRST symmetries for the
gauge- and (anti-)ghost fields of the theory as
\footnote{For the non-Abelian gauge theory where $F^{(2)} = d A^{(1)}
+ A^{(1)} \wedge A^{(1)}$, the off-shell nilpotent symmetry transformations
for the gauge (i.e. $s_b A_\mu = D_\mu C$)
and (anti-)ghost fields (with $s_b C = \frac{1}{2} C \times C$, etc.)
were found in a beautiful
paper by {\bf Bonora and Tonin}
with exactly the same kind of expansion as given in (8)
(see, [6] for details). The horizontality condition
($\tilde F^{(2)} = F^{(2)}$) plays an important
role in this case, too.}
$$
\begin{array}{lcl}
B^{(h)}_{\mu}\; (x, \theta, \bar \theta) &=& A_{\mu} (x)
+ \;\theta\; (s_{ab} A_{\mu} (x))
+ \;\bar \theta\; (s_{b} A_{\mu} (x))
+ \;\theta \;\bar \theta \;(s_{b} s_{ab} A_{\mu} (x)), \nonumber\\
{\cal F}^{(h)}\; (x, \theta, \bar \theta)
&=& C (x) \;+ \; \theta\; (s_{ab} C (x))
\;+ \;\bar \theta\; (s_{b} C (x))
\;+ \;\theta \;\bar \theta \;(s_{b}\; s_{ab} C (x)),
 \nonumber\\
\bar {\cal F}^{(h)}\; (x, \theta, \bar \theta) &=& \bar C (x)
\;+ \;\theta\;(s_{ab} \bar C (x)) \;+\bar \theta\; (s_{b} \bar C (x))
\;+\;\theta\;\bar \theta \;(s_{b} \;s_{ab} \bar C (x)).
\end{array} \eqno(8)
$$
The above exercise provides  the physical interpretation for the
(anti-)BRST charges $Q_{(a)b}$
as simply the generators (cf. (3)) of translations
(i.e. $ \mbox{Lim}_{\bar\theta \rightarrow 0} (\partial/\partial \theta),
 \mbox{Lim}_{\theta \rightarrow 0} (\partial/\partial \bar\theta)$)
along the Grassmannian directions of the supermanifold. It is obvious that
now $\tilde d \tilde A^{(1)}_{(h)} = d A^{(1)}$, where
$\tilde A^{(1)}_{(h)} = dx^\mu B_\mu^{(h)} + d \theta \bar {\cal F}^{(h)}
+ d \bar \theta {\cal F}^{(h)}$ is the modified version of the 1-form super
connection $\tilde A^{(1)}$ (cf. (6)) after
the application of the horizontality (soul-flatness) condition.

We now derive the nilpotent symmetry transformations
for the matter (Dirac) fields $(\psi, \bar\psi)$ due to the
invariance of the conserved matter current of the theory
on the supermanifold. We start off with the super expansion of the
superfields $(\Psi, \bar\Psi)(x, \theta,\bar\theta)$),
corresponding to the ordinary Dirac fields $(\psi, \bar\psi)(x)$ of
the Lagrangian density (1),
as [10,12]
$$
\begin{array}{lcl}
 \Psi (x, \theta, \bar\theta) &=& \psi (x)
+ i \;\theta\; \bar b_1 (x) + i \;\bar \theta \; b_2 (x)
+ i \;\theta \;\bar \theta \;f (x),
\nonumber\\
\bar \Psi (x, \theta, \bar\theta) &=& \bar \psi (x)
+ i\; \theta \;\bar b_2 (x) + i \;\bar \theta \; b_1 (x)
+ i\; \theta \;\bar \theta \;\bar f (x).
\end{array} \eqno(9)
$$
In the limit $(\theta, \bar\theta) \rightarrow 0$, from the above expansions,
we get back the usual Dirac fields $(\psi, \bar\psi)$ (of
the Lagrangian density (1)) and the number of
bosonic fields ($b_1, \bar b_1, b_2, \bar b_2)$ match with the fermionic
fields $(\psi, \bar \psi, f, \bar f)$ for the consistency
with supersymmetry.

We construct the
supercurrent $\tilde J_\mu (x, \theta, \bar\theta)$ with the
following general super expansion
$$
\begin{array}{lcl}
\tilde J_\mu (x, \theta, \bar\theta) = \bar \Psi (x,\theta,\bar\theta)
\;\gamma_\mu \;\Psi (x, \theta, \bar\theta)
= J_\mu (x) + \theta \; \bar K_\mu (x)
+ \bar \theta\; K_\mu (x) + i \; \theta\; \bar\theta\; L_\mu (x),
\end{array} \eqno(10)
$$
where the above components (i.e. $\bar K_\mu, K_\mu, L_\mu, J_\mu$),
can be expressed in terms of the components of the
basic super expansions (9), as (see, e.g., [10])
$$
\begin{array}{lcl}
&& \bar K_\mu (x) = i \bigl ( \bar b_2 \gamma_\mu \psi -
\bar \psi \gamma_\mu \bar b_1 \bigr ),
\qquad  K_\mu (x) = i \bigl ( b_1 \gamma_\mu \psi -
\bar \psi \gamma_\mu  b_2 \bigr ), \nonumber\\
&& L_\mu (x) = \bar f \gamma_\mu \psi + \bar \psi \gamma_\mu f
+ i (\bar b_2 \gamma_\mu b_2 - b_1 \gamma_\mu \bar b_1),
\qquad J_\mu (x) = \bar \psi  \gamma_\mu \psi.
 \end{array} \eqno(11)
$$
To be consistent with our earlier observation that the (anti-)BRST
transformations $(s_{(a)b})$ are equivalent to the translations
along the $(\theta)\bar\theta$-directions of the supermanifold,
it is straightforward
to re-express the expansion in (10) as
$$
\begin{array}{lcl}
\tilde J_\mu (x, \theta, \bar\theta) = J_\mu (x) + \theta \;
(s_{ab} J_\mu (x)) + \bar \theta\; (s_b J_\mu (x))
+ \theta\; \bar\theta\; (s_b s_{ab} J_\mu (x)).
\end{array} \eqno(12)
$$
It can be checked explicitly that, under the (anti-)BRST transformations (2),
the conserved current $J_\mu (x)$ remains invariant
(i.e. $s_{b} J_\mu (x) = s_{ab} J_\mu (x) = 0$). Thus, from (11), we have
$$
\begin{array}{lcl}
b_1 \gamma_\mu \psi
= \bar \psi \gamma_\mu b_2, \qquad
\bar b_2 \gamma_\mu \psi
= \bar \psi \gamma_\mu \bar b_1, \qquad
\bar f \gamma_\mu \psi
+ \bar \psi \gamma_\mu f = i (b_1 \gamma_\mu \bar b_1 - \bar b_2
\gamma_\mu b_2),
\end{array} \eqno(13)
$$
as the conditions for $s_{(a)b} J_\mu = 0$. This, ultimately,
implies:  $K_\mu = L_\mu = \bar K_\mu = 0$ in (10).

One of the possible solutions to the above restrictions, present in (13),
is [10]
$$
\begin{array}{lcl}
&& b_1 = - e \bar \psi C, \qquad b_2 = - e C \psi,
\qquad \bar b_1 = - e \bar C \psi, \qquad \bar b_2 = - e \bar \psi \bar C,
\nonumber\\
&& f = - i e\; [\; B + e \bar C C\; ]\; \psi,
\qquad \bar f = + i e\; \bar \psi\; [\; B + e C \bar C \;].
\end{array} \eqno(14)
$$
It is evident that the above expressions are consistent but {\it not} uniquely
determined by the restriction
$\tilde J_\mu (x, \theta, \bar\theta) = J_\mu (x)$ on the supermanifold.
However, it should be emphasized that, barring the constant factors,
the above solutions are very logical. For instance, for the
validity of $ b_1 \gamma_\mu \psi = \bar \psi \gamma_\mu b_2$, the pair
of bosonic fields $b_1$ and $b_2$ should be proportional to the fermionic
fields $\bar \psi$ and $\psi$, respectively. The
corresponding equality can be achieved,
{\it only} by bringing in, the (anti-)ghost fields of the theory. There
is {\it no} other possible choice. Thus, we judiciously choose
$b_1 \sim \bar\psi C$ and $b_2 \sim C \psi$. Rest of the consistent choices of
(14) are made on similar line of arguments with
appropriate constants $i$ and $e$ thrown in.

The stage is now set for the exact derivation of (14). To this end in mind,
we begin with the following {\it gauge invariant} restriction  on the
supermanifold [12]
$$
\begin{array}{lcl}
\bar \Psi (x, \theta,\bar\theta)\; \bigl (\tilde d
+ i e \tilde A^{(1)}_{(h)}\bigr )\;
\Psi (x, \theta, \bar\theta) = \bar \psi (x)\; (d + i e A^{(1)})\; \psi (x),
\end{array}\eqno(15)
$$
where the superfields $\Psi$ and $\bar \Psi$ are from (9). The r.h.s. of the
above equation, expressed  in terms of the differential $dx^\mu$ (as
$dx^\mu \bar \psi (\partial_\mu + i e A_\mu) \psi$),  is
obviously a $U(1)$ gauge invariant quantity. The l.h.s. of the above equation
yields the coefficients of the differentials $dx^\mu, d \theta$
and $d \bar\theta$. The analogues of the latter two,
as is evident from (15),  do not exist on the r.h.s.

It is straightforward to note that the coefficients of $d\theta$, collected
from the l.h.s., should be set equal to zero. This requirement
leads to the following two independent relationships
$$
\begin{array}{lcl}
- i\; \bar \psi\; \bigl (\bar b_1 + e \bar C \psi \bigr ) = 0, \qquad
\bar \psi\; \bigl (i f + e \bar C  b_2 - e B \psi \bigr ) = 0.
\end{array}\eqno(16)
$$
Similarly, the coefficients of $d \bar\theta$ equal to zero, implies the
following relationships [12]
$$
\begin{array}{lcl}
- i\; \bar \psi\; \bigl (b_2 + e  C \psi \bigr ) = 0, \qquad
\bar \psi\; \bigl  (- i f + e  C \bar b_1 + e B \psi \bigr ) = 0.
\end{array}\eqno(17)
$$
Together, the above two equations,
lead to the following results
(for $\bar \psi \neq 0$)
$$
\begin{array}{lcl}
\bar b_1 = - e \; \bar C\; \psi, \qquad
b_2 = - e\; C\;\psi, \qquad
f = - i e \;\bigl (B + e \bar C C\bigr )\; \psi.
\end{array}\eqno(18)
$$
In fact, out of {\it exactly} four relations, only
{\it two} in (16) and (17), are independent [12].

We shall focus now on the collection of the
coefficients of $dx^\mu, dx^\mu (\theta),
dx^\mu (\bar\theta)$ and $dx^\mu (\theta\bar\theta)$. The coefficient of
the ``pure'' $dx^\mu$ match from the l.h.s. and r.h.s. Exploiting the inputs
from (18), we set equal to zero the coefficient of $dx^\mu (\theta)$ and
$dx^\mu (\bar\theta)$. These imply
$$
\begin{array}{lcl}
i\;[\;\bar b_2 + e \; \bar \psi \bar C\;]\; [\;D_\mu \psi\;] = 0, \qquad
i\;[\;b_1 + e \; \bar \psi C\;]\; [\;D_\mu \psi\;] = 0.
\end{array}\eqno(19)
$$
The above conditions lead to the exact determination of $b_1$ and
$\bar b_2$ as: $ b_1 = - e \bar \psi C, \bar b_2 = - e \bar \psi \bar C$.
Here, it will be noted that $D_\mu \psi \neq 0$ for
the QED with Dirac fields. Finally, we collect the coefficients of
$dx^\mu (\theta\bar\theta)$
and set them equal to zero. This condition implies [12]
$$
\begin{array}{lcl}
[\; i \bar f + e \bar \psi \; (B + e C \bar C)\; ]\; [\; D_\mu \psi \;] = 0,
\end{array}\eqno(20)
$$
where we have exploited the inputs from (18) and have inserted the
values of $b_1$ and $\bar b_2$ that were obtained earlier. It is
obvious that, for $D_\mu \psi \neq 0$, we obtain the exact value
of $\bar f$ as: $\bar f = i e [B + e C \bar C] \bar \psi$. Thus,
from the restriction (15), we obtain {\it exactly} all the values of (14).
Insertions of the values of (14) into (9) leads to the following
(see, [12] for details)
$$
\begin{array}{lcl}
\Psi\; (x, \theta, \bar \theta) &=& \psi (x) \;+ \; \theta\;
(s_{ab}  \psi (x))
\;+ \;\bar \theta\; (s_{b} \psi (x))
\;+ \;\theta \;\bar \theta \;(s_{b}\;  s_{ab} \psi (x)),
 \nonumber\\
\bar \Psi\; (x, \theta, \bar \theta) &=& \bar \psi (x)
\;+ \;\theta\;(s_{ab} \bar \psi (x)) \;+\bar \theta\; (s_{b} \bar \psi (x))
\;+\;\theta\;\bar \theta \;(s_{b} \; s_{ab} \bar \psi (x)).
\end{array} \eqno(21)
$$
This establishes the fact that the nilpotent (anti-)BRST charges $Q_{(a)b}$
are the translations generators
$(\mbox{Lim}_{\bar \theta \rightarrow 0}(\partial/\partial \theta))
\mbox{ Lim}_{ \theta \rightarrow 0}(\partial/\partial \bar \theta)$
along the $(\theta)\bar\theta$ directions
of the supermanifold.

To summarize,
 the geometrical interpretations for (i) the
(anti-)BRST transformations $s_{(a)b}$ and their corresponding generators
$Q_{(a)b}$, (ii) the nilpotency property of $s_{(a)b}$ and $Q_{(a)b}$, and
(iii) the anticommutativity property of $s_{(a)b}$ and $Q_{(a)b}$, for
{\it all} the fields of QED with Dirac fields, emerge in the framework of
augmented superfield formalism. Mathematically, these can be expressed,
in an explicit manner, as illustrated below
$$
\begin{array}{lcl}
&& s_{b}\; \Leftrightarrow\; Q_{b}\; \Leftrightarrow\;
\mbox{Lim}_{\theta \rightarrow 0}
{\displaystyle \frac{\partial}{\partial
\bar\theta}}, \;\qquad\;
s_{ab}\; \Leftrightarrow\; Q_{ab} \;\Leftrightarrow\;
\mbox{Lim}_{\bar\theta \rightarrow 0}
{\displaystyle \frac{\partial}{\partial
\theta}}, \nonumber\\
&& s_{(a)b}^2 = 0\;\; \Leftrightarrow \;\;Q_{(a)b}^2 = 0\;\; \Leftrightarrow\;
\;\Bigl ({\displaystyle \frac{\partial}{\partial
\theta}} \Bigr )^2 = 0, \;
\Bigl ({\displaystyle \frac{\partial}{\partial
\bar\theta}} \Bigr )^2 = 0,\nonumber\\
&& s_b s_{ab} + s_{ab} s_b = 0\; \Leftrightarrow\; Q_b Q_{ab} + Q_{ab} Q_b = 0
\;\Leftrightarrow \;
{\displaystyle \frac{\partial}{\partial \bar\theta}
\frac{\partial}{\partial \theta} +
\frac{\partial}{\partial \theta}
\frac{\partial}{\partial \bar \theta}} = 0.
\end{array}\eqno(22)
$$
The {\it exact} nilpotent (anti-)BRST symmetries for the matter
(Dirac) fields are obtained from the gauge {\it invariant} restriction (15) on
the supermanifold which is different in nature than the gauge {\it covariant}
restriction of the horizontality condition (5) (see, [13] for details).

\end{document}